\documentclass[12pt]{article}
\usepackage[colorlinks=true, urlcolor=navyblue, linkcolor=navyblue, citecolor=navyblue]{hyperref}
\usepackage{graphicx,amsmath,amssymb}
\usepackage{indentfirst}
\usepackage[usenames]{color}
\usepackage{epstopdf}
\usepackage{enumerate}
\usepackage{appendix}
\usepackage{lineno}
\usepackage{setspace}
\usepackage{ulem}

\newcommand{\DZero}{\rm D\O}

\newcommand{\redt}{\textcolor{red}}
\newcommand{\bluet}{\textcolor{blue}}

\definecolor{navyblue}{rgb}{0,0.08,0.45}
\definecolor{darkred}{rgb}{0.7,0.0,0.0}
\definecolor{darkgreen}{rgb}{0,0.6,0.2}

\newcommand{\beq}{\begin{equation}}
\newcommand{\enq}{\end{equation}}
\newcommand{\beqa}{\begin{eqnarray}}
\newcommand{\beqast}{\begin{eqnarray*}}
\newcommand{\enqa}{\end{eqnarray}}
\newcommand{\enqast}{\end{eqnarray*}}
\newcommand{\nn}{\nonumber}

\newcommand{\bec}{\begin{center}}
\newcommand{\enc}{\end{center}}
\newcommand{\beqo}{\begin{quote}}
\newcommand{\enqo}{\end{quote}}
\newcommand{\bem}{\begin{minipage}}
\newcommand{\enm}{\end{minipage}}

\newcommand{\req}[1]{(\ref{#1})}

\newcommand{\half}{\textstyle \frac{1}{2}}

\newcommand{\ze}{\zeta}

\newcommand{\la}{\lambda}

\newcommand{\si}{\sigma}

\newcommand{\vp}{\varphi}

\newcommand{\De}{\Delta}

\definecolor{green}{rgb}{0,.5,0}

\begin{document}



\begin{center}

{\huge }

\vspace{6pt}

{\huge  Hadronic Superpartners from Superconformal and Supersymmetric Algebra}
\end{center}

\vspace{10pt}

\centerline{Marina Nielsen\footnote{mnielsen@if.usp.br}}

\vspace{3pt}

\centerline{\it Instituto de F\'isica, Universidade de S\~ao Paulo}

  \centerline{\it Rua do Mat\~ao, Travessa R187, 05508-090 S\~ao 
Paulo, S\~ao Paulo, Brazil}
\centerline {\it SLAC National Accelerator Laboratory, Stanford University,
  Stanford, CA 94309, USA}

\vspace{10pt}

\centerline{Stanley J. Brodsky\footnote{sjbth@slac.stanford.edu}}

\vspace{3pt}

\centerline {\it SLAC National Accelerator Laboratory, Stanford University,
  Stanford, CA 94309, USA}

\vspace{10pt}


\begin{abstract}

\vspace{15pt}
Through the embedding of superconformal quantum mechanics into AdS space, it is possible to construct an effective supersymmetric  QCD  light-front Hamiltonian for hadrons, which includes a spin-spin interaction between the hadronic constituents. A specific breaking of conformal symmetry determines a unique effective quark-confining potential for light hadrons, as well as remarkable connections between the meson, baryon, and tetraquark spectra. The pion is massless in the chiral limit and has no supersymmetric partner. The excitation spectra of relativistic light-quark meson, baryon and tetraquark bound states lie on linear Regge trajectories with identical slopes in the radial and orbital quantum numbers. Although conformal symmetry is strongly broken by the heavy quark mass, the basic underlying supersymmetric mechanism, which transforms mesons to baryons (and baryons to tetraquarks) into each other, still holds and gives remarkable connections across the entire spectrum of light, heavy-light and double-heavy hadrons. Here we show that all the observed hadrons can be related through this effective supersymmetric  QCD, and that it can be used to identify the structure of the new charmonium states.
\end{abstract}


\section{Introduction \label{introduction}}

Superconformal algebra allows  the construction of  relativistic light-front  (LF) semiclassical bound-state equations in physical spacetime which can be embedded in a higher dimensional classical gravitational theory. In a series of recent articles~\cite{deTeramond:2014asa, Dosch:2015nwa, Dosch:2015bca, Brodsky:2016yod,Dosch:2016zdv}, it was shown that this new approach to hadron physics includes the emergence of a mass scale and color confinement out of a classically scale-invariant theory, the occurrence of a zero-mass bound state, universal Regge trajectories for both mesons and baryons,  and the breaking of chiral symmetry in the hadron spectrum. This holographic approach to hadronic physics gives remarkable connections between the light meson and nucleon spectra~\cite{Dosch:2015nwa}, and even though heavy quark masses break conformal invariance, an underlying dynamical supersymmetry still holds in the light-heavy
sector~\cite{Dosch:2015bca,Dosch:2016zdv}.

 The emerging dynamical supersymmetry between mesons and baryons in this framework is not a consequence of supersymmetric QCD at the level of fundamental fields, but it represents the  supersymmetry between the LF bound-state wave functions of mesons and baryons. This symmetry is consistent with an essential feature of color $SU(N_C)$:  a cluster of $N_C-1$ constituents  can be in the same color representation as the anti-constituent; for $SU(N_c=3)$ this means {$\bf \bar 3_c \in  \bf 3_c \times \bf 3_c$ and $\bf  3_c \in  \bf \bar3_c \times \bf \bar3_c$}. This was the basis of early attempts~\cite{Miyazawa:1966mfa, Catto:1984wi, Lichtenberg:1999sc} to combine mesons and baryons in supermultiplets. A crucial feature of this formalism is that the supermultiplets consist of a meson wave function with internal LF angular momentum  $L_M$ and the corresponding baryon wave function with angular momentum $L_B = L_M - 1$ with the same mass. The $L_M = 0$ meson has no supersymmetric partner.

 In   AdS$_5$  the positive and negative-chirality projections of the baryon wave functions, the upper and lower spinor components in the chiral representation of Dirac matrices, satisfy uncoupled second-order differential equations with degenerate eigenvalues.  In particular, it was shown in \cite{Brodsky:2014yha} that the nucleon wave-function has equal probablility for states with $L=0$ and $L=1$, which implies that the spin $S_z$ of the quark in the proton has equal probability to be aligned or anti-aligned with the proton total spin $J_z$. These component wave functions form, together with the meson wave functions,  the supersymmetric multiplets. The model also predict the existence of tetraquarks which are degenerate with baryons with the same angular momentum. The tetraquarks are bound states of the same confined color-triplet diquark and anti-diquark clusters which account for baryon spectroscopy; they are required to complete the 4-supermultiplet structure  predicted by the superconformal algebra \cite{Brodsky:2016yod}. 
We emphasize that the supersymmetric relations which are derived from supersymmetric quantum mechanics are not based on a supersymmetric Lagrangian in which QCD is embedded; instead, they are based on the fact that the supercharges of the supersymmetric algebra relate the masses and the wave functions of mesons to baryons, and of baryons to tetraquarks, in a Hilbert space in which the LF Hamiltonian acts.  The properties of the supercharges predict specific constraints between mesonic and baryonic superpartners in agreement with measurements across the entire hadronic spectrum.

\begin{figure}[h]
\setlength{\unitlength}{1mm}
\begin{center}
\begin{picture}(60,30)(15,0)
\put(20,30){\circle{20}} \put(20,33){\circle*{2}}
\put(20,27){\circle{2}}
 \put(50,30){\circle{20}}
\put(48,33){\circle{2}} \put(52,33){\circle{2}}
\put(50,27){\circle{2}}
 \put(14,19){$\phi_M,\;L_M$}
\put(43,19){$\psi_{B+},\;L_M-1$}
 \put(22,33){\vector(1,0) {24}}
\put(32,35){$Q^\dagger$}
\end{picture}
\end{center}
\vspace{-60pt}
\caption{\label{me-ba} \small Open circles represent quarks, full circles antiquarks. The baryon has the same mass as its meson partner in the multiplet. }
\end{figure}
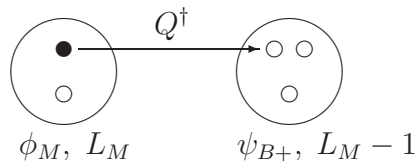

The supercharges operators, $Q^\dagger$, can be interpreted as transforming a constituent into a two-anti-constituent cluster in the same color representation as the constituent. It transforms a quark into an anti-diquark in color representation ${\bf 3_c}$ and an antiquark into a diquark in color representation ${\bf\bar 3_c}$.   Therefore  the operator $Q^\dagger$ applied to the meson wave function with internal LF angular momentum  $L_M$ will give the positive-chirality component of  the corresponding baryon wave function with angular momentum $L_B = L_M - 1$ with the same mass,  as depicted in Fig.~\ref{me-ba}.

\vspace{30pt}

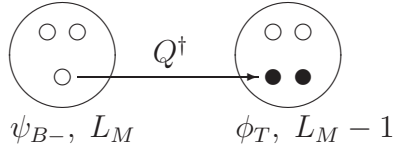
\begin{figure}[h]
\setlength{\unitlength}{1mm}
\begin{center}
\begin{picture}(60,30)(15,0)
 \put(50,7){\circle{20}}
\put(48,10){\circle{2}} \put(52,10){\circle{2}}
\put(50,4){\circle{2}} \put(43,-4){$\psi_{B-}, \;L_M$}
\put(80,7){\circle{20}} \put(78,10){\circle{2}}
\put(82,10){\circle{2}} \put(78,4){\circle*{2}}
\put(82,4){\circle*{2}} \put(73,-4){$\phi_T,\;L_M-1$}
\put(52,4){\vector(1,0) {24}} \put(62,6){$Q^\dagger$}
\end{picture}
\end{center}
\caption{\label{tetra} \small Open circles represent quarks, full circles antiquarks.  The tetraquark has the same mass as its baryon partner in the multiplet. }
\end{figure}

On the other hand, the negative-chirality component of a baryon, $\psi_{B-}$, has LF angular momentum $L_B+1$ if its positive-chirality component partner has LF angular momentum $L_B$.  Since $Q^\dagger$ lowers the angular momentum by one unit, the angular momentum of the corresponding tetraquark is  equal to that of the positive-chirality component of the baryon, $L_T=L_B=L_M-1$. Therefore,  the operator $Q^\dagger$ applied to the negative-chirality component of a baryon will give a tetraquark wave function representing a bound state of a diquark and an anti-diquark cluster,  as depicted in Fig.~\ref{tetra}.

From Figs.~\ref{me-ba} and ~\ref{tetra} we see that the fundamental structure of a baryon is a quark-diquark; for a tetraquark the fundamental structure is a diquark-antidiquark. It is important to mention that diquarks have been very well studied in the past
\cite{Shuryak:2005pk,Wilczek:2004im,Selem:2006nd,Alexandrou:2005zn}
and are very useful degrees of freedom to focus on in QCD. In this superconformal holographic QCD approach, the diquarks are themselves composite and thus not pointlike. They are rather a consequence of the light-front cluster decomposition and, as we are going to show, they help to organize all the known hadronic spectra.

This article is organized as follows: in Sec.~2 we give a brief review of the superconformal algebra approach. In Sec.~3 we relate the meson-baryon-tetraquark superpartners. Finally, in Sec.~4 we present our conclusions.

\section{Supersymmetric quantum mechanics and hadron physics}

We briefly review the essential features of the supersymmetric quantum mechanics
 and superconformal algebra \cite{Witten:1981nf,Fubini,deAlfaro:1976vlx,Haag} as aplied to the LF Hamiltonian \cite{Dosch:2015nwa, Dosch:2015bca, Brodsky:2016yod,Dosch:2016zdv}. The Hamiltonian can be written in terms of two fermionic generators, the supercharges, $Q$ and $Q^\dagger$, which satisfy anticommutations relations. The Hamiltonian commutes with these fermionic generators:
\beq\label{HQ}
H=\{Q,Q^\dagger\},
\enq
\beq\label{QQ}
\{Q,Q\}=\{Q^\dagger,Q^\dagger\}=0,~[H,Q]=[H,Q^\dagger]=0.
\enq
In matrix notation the supercharges and the Hamiltonian can be written as 
\beq Q =
\left(\begin{array}{cc}
0&q\\
0&0\\
\end{array}
\right) ,\quad Q^\dagger=\left(\begin{array}{cc}
0&0\\
q^\dagger&0\\
\end{array}
\right) ,\quad
H= \left(\begin{array}{cc}
q \, q^\dagger &  0\\
0 & q^\dagger q \\
\end{array}
\right)  \label{QQH},
\enq with \beqa \label{qdag}
q &=&-\frac{d}{d \ze} + \frac{f}{\ze} + V(\ze),\\
q^\dagger &=& \frac{d}{d\ze}  + \frac{f}{\ze} + V(\ze), \enqa
where $\ze$ is the LF invariant transverse variable and  $f$ is a
dimensionless constant. For an arbitrary $V(\ze)$ the resulting Hamiltonian is no longer conformal, but it is supersymmetric, and this supersymmetric quantum mechanics was first proposed by Witten \cite{Witten:1981nf}. For
\beq
V(\ze)=\lambda~\ze
\enq
the supercharge $Q$ is a superposition of fermion generators inside the superconformal algebra \cite{Fubini}. The resulting supersymmetric superconformal light-front  Hamiltonian is \cite{Dosch:2015bca}:
\beq \label{h1}
 H
  =     \left(\begin{array}{cc} - \frac{d^2}{d \ze^2}+\frac{4 L_M^2-1}{4\ze^2}+U_M(\ze)& \hspace{-1cm} 0 \\
0 & \hspace{-1cm} - \frac{d^2}{d \ze ^2} +\frac{4 L_B^2-1}{4\ze
^2}+U_B(\ze )
\end{array}\right),
  \enq
where the effective LF potentials of both, mesons and baryons, are:
\beqa
 U_M(\ze) & = &  \la^2 \ze^2 + 2 \, \la (L_M - 1),   \label{UM} \\
 U_B(\ze) & = &  \la^2 \ze^2 +  2\, \la (L_B + 1).  \label{UB}
 \enqa
with $L_B + \half = L_M - \half \;=f$, from which follows the crucial relation $L_M=L_B+1$.   These quadratic confining potencials are holographically related to a unique dilaton profile, $\varphi(\ze)= +\lambda \zeta^2$ \cite{Brodsky:2014yha}. 

 Althought the eigenstates of both operators in $H$ are different, they have the same supersymmetric eigenvalues \cite{Brodsky:2014yha}:
 \beq
 M^2_{n,f}=4\lambda\left(n+f+{1\over2}\right),
 \enq
where $n$ is the radial quantum number. 
 
 One can explicitly break conformal symmetry  without violating supersymmetry by adding  to the Hamiltonian  \req{h1} a multiple of the unit matrix, $\mu^2 {\bf I}$:
\beq
\label{hmmu} H_\mu =
\left(\begin{array}{cc} - \frac{d^2}{d \ze^2}+\frac{4 L_M^2-1}{4\ze^2}+U_M(\ze)& \hspace{-1cm} 0 \\
0 & \hspace{-1cm} - \frac{d^2}{d \ze ^2} +\frac{4 L_B^2-1}{4\ze
^2}+U_B(\ze )
\end{array}\right) + \mu^2 {\bf I},
\enq 
where the constant   term   $\mu^2$  contains the effects of spin coupling and
quark masses. This term has been derived for light hadrons in
Ref.~\cite{Brodsky:2016yod}, yielding very satisfactory results. In particular, in Ref.~\cite{Brodsky:2016yod} it was shown that the effects of spin coupling can be separated from the quark masses correction in the following form:
\beq
\mu^2\,\to\, 2 \la S\,+\,\Delta M^2[m_1,...,m_N],
\enq
where $S$ is the internal quark spin for mesons and the lowest possible value of the diquark cluster spin for baryons. $\Delta M^2[m_1,...,m_N]$ is the quark masses correction which expression is given in Ref.~\cite{Brodsky:2016yod}. Since the effect of the term $\mu^2{\bf I}$ in the new Hamiltonian is an overall shift of the mass scale, it does not change the LF wavefunction.

\vspace{20pt}
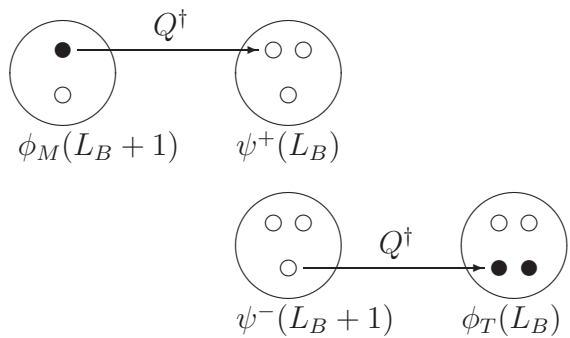
\begin{figure}[h]
\setlength{\unitlength}{1mm}
\begin{center}
\begin{picture}(60,30)(15,0)
\put(20,30){\circle{20}} \put(20,33){\circle*{2}}
\put(20,27){\circle{2}}
 \put(50,30){\circle{20}}
\put(48,33){\circle{2}} \put(52,33){\circle{2}}
\put(50,27){\circle{2}}
 \put(14,19){$\phi_M(L_B+1)$}
\put(43,19){$\psi^{+}(L_B)$}
 \put(22,33){\vector(1,0) {24}}
\put(32,35){$Q^\dagger$}

 \put(50,7){\circle{20}}
\put(48,10){\circle{2}} \put(52,10){\circle{2}}
\put(50,4){\circle{2}} \put(43,-4){$\psi^{-}(L_B+1)$}
\put(80,7){\circle{20}} \put(78,10){\circle{2}}
\put(82,10){\circle{2}} \put(78,4){\circle*{2}}
\put(82,4){\circle*{2}} \put(73,-4){$\phi_T(L_B)$}
\put(52,4){\vector(1,0) {24}} \put(62,6){$Q^\dagger$}
\end{picture}
\end{center}
\caption{\label{tetra2} \small The supersymmetric 4-plet 
$\{\phi_M, \psi_{B+}, \psi_{B-},\phi_T\}$. Open and full circles stand for
quarks and antiquarks respectively. }
\end{figure}

As shown in \cite{Brodsky:2016yod}, the full supersymmetric approach described above leads to a 4-plet,  as ilustrated in Fig.~\ref{tetra2}.
As mentioned in the Introduction, the operator $Q^\dagger$ can be interpreted as transforming a constituent into a two-anti-constituent cluster in the same color representation as the constituent. It transforms a quark into an anti-diquark in color representation ${\bf 3_c}$ and an antiquark into a diquark in color representation ${\bf\bar 3_c}$.  Since $Q^\dagger$ lowers the angular momentum by one unit, the operator $Q^\dagger$ applied to the meson wave function, $\phi_M$, with angular momentum $L_M$ leads to the leading-twist positive chirality component of a baryon, $\psi^+$, with angular momentum $L_B=L_M -1$ ~\cite{deTeramond:2014asa, Brodsky:2014yha}. On the other hand, the operator $Q^\dagger$ applied to the negative-chirality component of a baryon, $\psi^-$, with angular momentum ~~$L_B+1$ will give a tetraquark wave function, a bound state of a diquark and an anti-diquark cluster, with angular momentum $L_T=L_B$ ~\cite{Brodsky:2016yod}.

The full supersymmetric quadruplet representation thus consists of two fermion wave functions, the positive and negative-chirality components of the baryon spinor wave function $\psi^{+}$ and $\psi^{-}$, plus two bosonic wave functions, the meson $\phi_M$ and the tetraquark $\phi_T$. These states can be arranged as a $2 \times2$ matrix~\cite{Brodsky:2016yod}:
\beq\label{multi}
\left(\begin{array}{cc}
\phi_M{(L_M = L_B+1)} &\psi^{-}{(L_B+ 1)}\\
\psi^{+}{(L_B)} &\phi_T{(L_T = L_B)}
\end{array}
\right), 
\enq
 on which the Hamiltonian~\req{hmmu} operate. The resulting expressions for the squared masses of the mesons, baryons and  tetraquarks are~\cite{Brodsky:2016yod}:
\beqa \label{mesfin}  
\mbox{Mesons:} && M_M^2 = 4 \la (n+L_M+{S_M\over2})+ \De M^2[m_1,m_2] ,\\
\label{barfin}\mbox{Baryons:} && M_B^2=4 \la (n+L_B+{S_D\over2}+1) + \De M^2[m_1,m_2,m_3] ,\\
\label{tetrafin}   \mbox{Tetraquarks:} && M_T^2=4 \la (n+L_T+{S_T\over2}+1) + \De M^2[m_1,m_2,m_3,m_4] ,
\enqa 
where $S_M$ is the meson spin, $S_D$ is the lowest possible value of the diquark cluster spin of the  baryons, while $S_T$ is the tetraquark spin.  The different values of the mass corrections, $\Delta M^2$, on the supermultiplet break supersymmetry explicitly. These equations show that the excitation spectra of relativistic light-quark meson, baryon and tetraquark bound states lie on linear Regge trajectories with identical slopes in the radial and orbital quantum numbers. Mesons with $L_M$ and $S_M$ are the superpartners of baryons with $L_B=L_M-1$ and the diquark with $S_D=S_M$. Analogously, baryons with $L_B$ and diquark with $S_D$  are the superpartners of tetraquarks with $L_T=L_B$, and $S_T=S_D$. 

It should be noted that there is a fundamental difference between the meson-baryon superpartners, versus the baryon-tetraquark superpartners which appear in the 4-plet representation of superconformal algebra. The lightest meson eigenstate, with $L_M=0$, does not have a baryonic superpartner, whereas the baryon ground state with $L_B=0$, does have a tetraquark superpartner.   The reason for this is related to the difference between the composition of the light-front wave functions of the bosonic and fermionic eigenstates. From Eqs.~\req{hmmu} and \req{multi} one can see that there is only one Light-Front Schr\"odinger equation for mesons:
  \beq
  (H_{\mu})_{11}\Phi_M(L_M)=M_M^2\Phi_M(L_M).
  \enq
However, there are two  Schr\"odinger equations for baryons, one for each
chirality projection of the baryon wave function:
  \beq
(H_\mu)_{11}\psi^-(L_B+1)=M_B^2\psi^-(L_B+1),\;\;(H_\mu)_{22}\psi^+(L_B)=M_B^2\psi^+(L_B).
  \enq
The baryon wave function has equal probability for the states components
with $L_B$ and $L_B+1$. When one says that a baryon has angular momentum $L_B$,
it refers to the angular momentum of the positive-chirality component of the
baryon wave function. Therefore, for a baryon with $L_B=0$, there will be
a tetraquark partner, also with $L_T=0$, since the tetraquark is related with the application of the operator $Q^\dagger$ into the negative-chirality component
of the baryon wave function with $L_B=1$.

In the presence of heavy quark masses the dilaton profile, $\varphi$, is not constrained by the superconformal algebraic structure and thus, in this case, its form and the form of the superpotential  $V$ are unknown {\it a priori}. However,
additional constraints  appear by  the holographic embedding.
In Ref.~\cite{Dosch:2016zdv} it was shown that the LF potential  in the heavy-light sector, even for strongly broken conformal invariance, has the same quadratic form as the one dictated by the conformal algebra:
 \beq
  \vp(\ze) = \half  \la A \,\ze^2 + B,  \quad \quad
  V(\ze) = \half \la A \, \ze, \label{phiV}
 \enq
 where the constant $A$ is indetermined. This means that the strength of the potential is not determined in the heavy-light sector. In Ref.~\cite{Dosch:2016zdv} it was also shown that the interaction potential is unchanged by adding a constant to the dilaton profile. Therefore,  the choice $B =0$ was made. This leads to the change $\la\to\la_Q$ in Eqs.~\req{mesfin},~\req{barfin},~\req{tetrafin}, where the slope constant, $\la_Q =\half  \la \, A$, can depend on the mass of the heavy quark \cite{Dosch:2016zdv}.

\section{Unraveling the Quark Structure of the Hadrons \label{qm}}

\subsection{Mesons and Baryons \label{ba-me}}

As shown in Ref.~\cite{Brodsky:2014yha}, the expression in Eq.~\req{mesfin} is valid only for mesons -  $q\bar{q}$ valence Fock states, with the maximum possible value for the total quantum number $J$, {\it i.e.}  $J=L_M+S_M$. In this LFHQCD approach, only mesons with $J=L_M+S_M$ will be considered as quark-antiquark states. For the other bosonic states, such as scalars, we are going to show that they can be identified as tetraquark states.

As mentioned above, baryons with angular momentum $L_B=L_M-1$ and diquark total spin $S_D=S_M$ are the superpartners of mesons with angular momentum $L_M$ and total spin $S_M$. From the relation $L_B=L_M-1$ we also see that mesons with $L_M=0$, as the $\pi$ and $\rho-\omega$, do not have a baryonic superpartner. The superpartners and the slopes that describes the Regge trajectories of the light quark mass mesons and baryons as well as mesons and baryons with strange, charm and bottom quarks are given in Refs.~\cite{Dosch:2015nwa,Dosch:2015bca,Dosch:2016zdv}.
{\renewcommand{\arraystretch}{1.1}  
\begin{table}[h]
\begin{center}
\begin{tabular}{| c|c | c|}
\hline
\multicolumn{2}{|c|}{State} & \\
I=1& I=0 & $L_M,~J^{PC}$ \\
\hline
$\pi(140)$ & & $0,~0^{-+}$\\
$b_1(1235)$& $h_1(1170)$&$1,~1^{+-}$\\
$\pi_2(1670)$&$\eta_2(1645)$&$2,~2^{-+}$\\
\hline
\end{tabular}
\end{center}
\caption{\small \label{pionfa}
  Quantum number assignment for the pion trajectory ($I=1$), and $h_1$  trajectory ($I=0$), corresponding to the leading-twist angular momentum. For a $q\bar{q}$ state $P=-(-1)^{L_M}$ and $C=(-1)^{L_M+S_M}$.}
\end{table}
As an example, let us consider the $\pi$  family consisting of states with  $n=0,~S_M=0,~I=1$ and $L_M=0,~1,~2,...$ given in Table~\ref{pionfa}. In this Table we have also included the isospin $I=0$ states  $h_1(1170)$ and $\eta_2(1645)$, since they are $q\bar{q}$ states with only $u$ and $d$ quarks. This is not the case of the isospin $I=0$ states $\eta(550)$ and $\eta^\prime(958)$, since for these states there is a strong mixing with $s\bar{s}$ constituents. Therefore, we include the $\eta(550)$ and $\eta^\prime(958)$ in the family of $s\bar{s}$ states with $n=0,~S_M=0$ and $L_M=0,~1,~2,...$ given in Table~\ref{etafa}.
{\renewcommand{\arraystretch}{1.1}  
\begin{table}[h]
\begin{center}
\begin{tabular}{| c | c|}
\hline
State $(I=0)$ & $L_M,~J^{PC}$ \\
\hline
$\eta(550),~\eta^\prime(958)$ & $0,~0^{-+}$\\
$h_1(1380)$&$1,~1^{+-}$\\
$\eta_2(1870)$&$2,~2^{-+}$\\
\hline
\end{tabular}
\end{center}
\caption{\small \label{etafa}
  Quantum number assignment for the $s\bar{s}$ states with $I=0$, $n=0,~S_M=0$ and $L_M=0,~1,~2,...$.} 
\end{table}

For bosonic clusters consisting of identical fermions, the relation beween spin and statistics gives specific constraints to the quantum numbers. Since the diquark cluster must be in a ${\bf\bar{3}_c}$ colour representation, it is antisymmetric in colour. This implies that the hadron must be both symmetric or both antisymmetric in spin and isospin. Hence, spin and isospin of the diquark cluster of light $u,~d$ quarks are related to the combinations $I_D = 1; S_D = 1$ or $I_D = 0; S_D = 0$.

The baryon superpartner of the $b_1(1235)$ and $h_1(1170)$ mesons has $n=0$, $L_B=L_M-1=0$ and a diquark cluster with total spin $S_D=S_M=0$. Therefore,  the isospin of the diquark cluster is also zero, which gives a baryon with quantum numbers $I~(J^P)={1\over2}~({1\over2}^+)$ (for a baryonic state $P=(-1)^{L_B}$). It is important to notice that both mesonic states, with $I=0$ and $I=1$, will be candidates for the superpartner of a baryon with $I=1/2$, if the diquark cluster has $S_D=0$. Hence, we identify the nucleon as the superpartner of the $b_1(1235)-h_1(1170)$ mesons. As pointed out in \cite{Dosch:2015nwa}, these states are somewhat heavier than its supersymmetric partner, the nucleon. However, it should be noted that the semiclassical equations of light-front holographic QCD and superconformal quantum mechanics used here are intended to be a zeroth order approximation to the complex problem of bound states in QCD. Therefore, we do not expect a perfect agreement with the data, but rather a global description of hadron spectroscopy that relates all possible supersymmetric partners.

In the case of the $\pi_2(1670)-\eta_2(1645)$ mesons, the only difference with the previous case is that these mesons have $L_M=2$. Therefore, the baryon superpartner has $L_B=L_M-1=1$ and quantum numbers $I~(J^P)={1\over2}~({1\over2}^-)$ and $I~(J^P)={1\over2}~({3\over2}^-)$. The candidates are the  first $L$ excitation of the nucleon: the $N^{3/2-}(1520)$ and $N^{1/2-}(1535)$ pair.

For the $s\bar{s}$ states in Table~\ref{etafa}, from the relation $L_B=L_M-1$ we  see that the states $\eta(550)$ and $\eta^\prime(958)$, with $L_M=0$,  do not have a baryonic superpartner. The superpartner of the $h_1(1380)$ is the $I~(J^P)={1\over2}~({1\over2})^+,~\Xi(1320)$.  Notice that for a $s\bar{s}$ mesonic state, the diquark cluster in the baryon superpartner has the quark content $sq$, where $q$ represents a $u$ or $d$ quark. Therefore, although the spin of the diquark is $S_D=0$, its isospin is $I_D=1/2$, as the isospin of the baryon. The superpartner of the $\eta_2(1870)$ meson is the first $L$ excitation  of $\Xi(1320)$. Although the quantum numbers of the states $\Xi(1620)$ and $\Xi(1690)$ have not yet been decisively determined, there is some indication that the $\Xi(1690)$ could be a $I~(J^P)={1\over2}~({1\over2}^-)$ state \cite{PDG}. We identify these two states as the first $L$ excitation of the $\Xi(1320)$ and, therefore, they are the natural candidates for the baryonic superpartners of the $\eta_2(1870)$ meson.

We can do a similar analysis for the $\rho-\omega$ family, namely the isovector and isoscalar states with $n=0,~S_M=1$ and $L_M=0,~1,~2,...$. This leads to baryonic superpartners with diquarks with total spin $S_D=1$ and, therefore, to the assignments given in Table~\ref{ro-de}.
{\renewcommand{\arraystretch}{1.1}  
\begin{table}[h]
\begin{center}
  \begin{tabular}{| c|c| c||c|c|}
\hline
\multicolumn{2}{|c|}{meson} &  & baryon & \\
I=1& I=0 &$L_M,~J^{PC}$ &$I=3/2$& $L_B,~J^P$\\
\hline
$\rho(770)$&$\omega(780)$& $0,~1^{--}$& --&\\
$a_2(1320)$&$f_2(1280)$&$1,~2^{++}$& $\De(1232)$& $0,~{3\over2}^+$\\
$\rho_3(1690)$&$\omega_3(1670)$&$2,~3^{--}$& $\De_{{3\over2}^-}(1700))$& $1,~{3\over2}^-$\\
$a_4(2040)$&$f_4(2050)$&$3,~4^{++}$& $\De_{{7\over2}^+}(1950))$& $2,~{7\over2}^+$\\
\hline
\end{tabular}
\end{center}
\caption{\small \label{ro-de}
  Quantum number assignment for the $\rho-\omega$ trajectory and baryonic superpartners. We show only the baryonic state with the higher possible value for $J$.}
\end{table}

\subsection{Tetraquarks \label{te}}

As pointed out in Ref.~\cite{Dosch-lec},  supersymmetry demands that only one of the clusters in the tetraquark can have $S_D = 1$, since otherwise a total quark spin 2 would be possible, which cannot be accomodated in the multiplet. Therefore, one of the clusters in the tetraquark, superpartner of the baryon, has always spin zero. Since the diquark cluster in the baryon has spin $S_D=S_M$, the operator $Q^\dagger$ transforms the quark, which is not in the diquark cluster in the baryon, into an anti-diquark cluster with spin zero. This ensures that the total spin of the tetraquark superpartner is  $S_T=S_D$ as required by supersymmetry (see Eqs.~\req{mesfin}, \req{barfin} and \req{tetrafin}).

In this paper we will use the notation introduced by Wilczek~\cite{Wilczek:2004im} to represent the diquarks. A diquark (or anti-diquark) with total spin $S_D=0$ and in the ${\bf \bar3_c}$ color representation was called a ``good-diquark'', and was represented by $[ud]$ (or $[\bar{u}\bar{d}]$). On the other hand, the author called a ``bad-diquark'' a diquark with total spin $S_D=1$ and represented it as $(ud)$ (or $(qq)$ in general).

\subsubsection{Light Quark States \label{te-qq}}

Tetraquark superpartners were assigned to the nucleon and to the $\Delta$, in Ref.~\cite{Brodsky:2016yod}, by noticing that the relative angular momentum of the two diquarks in the tetraquark, $L_T$, is equal to the angular momentum $L_B$ of the positive-chirality component of the baryon and, therefore, its parity is $(-1)^{L_B}$, since it consists of an even number of antiquarks. As the leading-twist component of the nucleon has $L_B=0, S_D=0$, its tetraquark superpartner consists of a diquark and a anti-diquark, both with $I_D=0$ and $S_D=0$, and hence $S_T=0$, and with $L_T=L_B=0$, {\it i.e.}, it is a scalar-isoscalar tetraquark consisting of two scalar-isoscalar quark clusters: $|[ud][\bar{u}\bar{d}]\rangle$. Since such tetraquark state coincides with its antiparticle it has positive charge conjugation $C=+$. Therefore, its quantum numbers are $J^{PC}=0^{++}$, as the scalar bosonic states. This is why what are normally identified as scalar mesons are here classified as tetraquark states. In contrast, a quark-antiquark scalar state with $J^{PC}=0^{++}$ would have $L_M=1, ~S_M=1$ and $J=0\neq L_M+S_M$. The additivity property, $J=L_M+S_M$, is a distinguishing characteristic between ordinary $q\bar{q}$ mesons versus tetraquark states.

The idea that the light scalar mesons could be diquark-antidiquark bound states was first proposed by Jaffe in 1977 \cite{jaffe}, and has later
been extrapolated to heavier sectors. Jaffe proposed that some states may be composed of two quarks and two antiquarks in a diquark-antidiquark structure: $|[qq][\bar{q}\bar{q}]\rangle$. In particular the quark content proposed in Ref. \cite{jaffe} for the  $f_0(980),\,a_0(980)$ and $\sigma(500)$ scalar states  are: 
\beqa\label{tetra-ja}
|\si\rangle=|[ud][\bar{u}\bar{d}]\rangle,\;\;\;\;\;\;&&|f_0\rangle={1\over\sqrt{2}}(|[us][\bar{u}\bar{s}]\rangle+|[ds][\bar{d}\bar{s}]\rangle),
\nn\\
|a_0^-\rangle=|[ds][\bar{u}\bar{s}]\rangle,\;\;\;\;\;\;&&|a_0^0\rangle={1\over\sqrt{2}}(|[us][\bar{u}\bar{s}]\rangle-|[ds][\bar{d}\bar{s}]\rangle),\;\;\;\;\;\;|a_0^+\rangle=|[us][\bar{d}\bar{s}]\rangle.
\enqa
In this four-quark scenario for the light scalars, the mass degeneracy of 
$f_0(980)$ and $a_0(980)$ is natural. 

As shown above, the tetraquark superpartner of the nucleon is a  scalar-isoscalar $|[ud][\bar{u}\bar{d}]\rangle$ state. Therefore,  the $\si(500)$ is a candidate, although the mass of the $f_0(980)$ is much closer to the nucleon mass than the $\si(500)$ mass. Probably, due to its huge width, $\Gamma_\sigma=(400-700)$ MeV, its nominal mass may not be taken as an indication of a big break in the supersymmety. Also, as pointed above, we do not expect a perfect agreement with the data, but rather a global picture classifying all possible supersymmetric partners.

Both $f_0(980)$ and $a_0(980)$ can be assigned as the tetraquark superpartners of the $\Sigma(1190)$. The $\Sigma(1190)$ is a baryonic state with quark structure $|[sq]q\rangle$ ($q$ represents a $u$ or $d$ quark). The diquark cluster, $[sq]$, has total spin $S_D=0$ and isospin $I_D={1\over2}$, so the $\Sigma(1190)$ has the same quantum numbers as the nucleon but isospin $I=1$. Since the supercharge $Q^\dagger$ transforms the constituent $q$ into an anti-diquark cluster, the anti-diquark $[\bar{q}\bar{s}]$, can be obtained leading to the tetraquark state $|[sq][\bar{s}\bar{q}]\rangle$ with total spin $S_T=0$ but isospin $I=1$ or $I=0$. Therefore, the states $a_0(980)$ and $f_0(980)$ are both candidates to be the tetraquark superpartners of the $\Sigma(1190)$.

The tetraquark superpartner of the $\De$ consists of a diquark with  $S_D=I_D=1$, and an anti-diquark, $[\bar{u}\bar{d}]$, with spin and isospin zero. The parity is $P=+$, since $L_T=L_B=0$. In this case, such a tetraquark: $|T_1\rangle=|(qq)[\bar{u}\bar{d}]\rangle$ is, in principle, distinguishable from its charge conjugate state,  $|\bar{T}_1\rangle=|(\bar{q}\bar{q})[{u}{d}]\rangle$, and, therefore, does not have a definite charge conjugation. As ilustrated in Fig.~4 , both states are obtained.

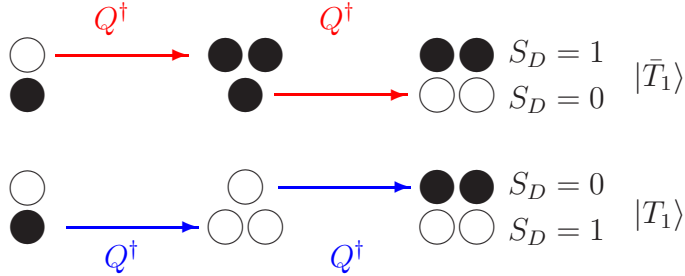
\begin{figure}
\label{scheme-tetra}
\begin{center}
\setlength{\unitlength}{0.5pt}
\begin{picture}(350,250)(0,-50)
\put(25,35){\circle*{25}}
\put(25,65){\circle{25}}
\put(25,-35){\circle{25}}
\put(25,-65){\circle*{25}}

\redt{\put(75,85){$Q^\dagger$}}
\bluet{\put(75,-95){$Q^\dagger$}}

\redt{\put(230,85){$Q^\dagger$}}
\bluet{\put(230,-95){$Q^\dagger$}}

{\put(365,60){$S_D=1$}} 
{\put(365,25){$S_D=0$}}
{\put(460,42){$|\bar{T}_1\rangle$}}
{\put(365,-40){$S_D=0$}} 
{\put(365,-75){$S_D=1$}} 
{\put(460,-62){$|{T}_1\rangle$}}

{\thicklines \redt{\put(15,65){\vector(1,0){100}}}}
{\thicklines \bluet{\put(15,-65){\vector(1,0){100}}}}

{\thicklines \redt{\put(165,35){\vector(1,0){100}}}}
{\thicklines \bluet{\put(160,-35){\vector(1,0){100}}}}

\put(120,65){\circle*{25}}
\put(150,65){\circle*{25}}
\put(135,35){\circle*{25}}

\put(120,-65){\circle{25}}
\put(150,-65){\circle{25}}
\put(135,-35){\circle{25}}

\put(280,65){\circle*{25}}
\put(310,65){\circle*{25}}
\put(280,35){\circle{25}}
\put(310,35){\circle{25}}

\put(280,-65){\circle{25}}
\put(310,-65){\circle{25}}
\put(280,-35){\circle*{25}}
\put(310,-35){\circle*{25}}

\end{picture}
\end{center}
\caption{ The  fermionic operator $Q^\dagger$ can be interpreted either as transforming a quark into an antiquark pair (red arrows) or an antiquark into a quark pair (blue arrows).}
\end{figure}

As pointed out in \cite{Dosch-lec}, states with definite charge conjugation can be obtained by the symmetric and antisymmetric superposition of these two tetraquarks:
\beqa \label{T1pm}
|T_1^+\rangle &=&{1\over\sqrt{2}}\left(|T_1\rangle+|\bar{T}_1\rangle\right),\;\;J^{PC}=1^{++},\\
|T_1^-\rangle &=&{1\over\sqrt{2}}\left(|T_1\rangle-|\bar{T}_1\rangle\right),\;\;J^{PC}=1^{+-}.
\enqa
A normal $1^{++}$ quark-antiquark state would have $L_M=1,~S_M=1$ and would not obey the additivity property ($J=L_M+S_M$). However, ${^{2S_M+L_M}(L_M)_J}={^1P_1}$ quark-antiquark mesonic states have $J^{PC}=1^{+-}$ and $J=1=L_M+S_M$, like the $b_1(1235)$ meson. Since states with same quantum numbers can mix, the $b_1(1235)$ is probably a mixture between quark-antiquark and tetraquark states. Therefore, here the  $b_1(1235)$ will not be identified as a mesonic neither as a tetraquark superpartner of a baryon. On the other hand, since the tetraquarks described above have $I=1$, the $h_1(1170)$ is a true quark-antiquark state and can, therefore, be considered as the mesonic superpartner of the nucleon, while the $a_1(1260)$ is a candidate for the tetraquark superpartner of the $\De$.

The first $L$ excitation of the nucleon is the $N^{3/2-}(1520)$ and $N^{1/2-}(1535)$ pair.  Its tetraquark superpartner should consists of two $I_D=0,  S_D=0$ clusters, and therefore, $C=+$,  with parity $P=-$, since $L_T=L_B=1$, and  quantum numbers $I~(J^{PC})=0~(1^{-+})$. The only states in \cite{PDG} with $1^{-+}$ quantum numbers are  the $\pi_1(1400)$ and $\pi_1(1600)$ both with $I=1$.  Since the $I=0,~L_T=0$, $|[ud][\bar{u}\bar{d}]\rangle$, $\sigma(500)$ state has already a huge width, its $L_T=1$ excitation would be broader or even unbound. Therefore, here we do not predict the existence of such a state.

The tetraquark superpartner of the first $L$ excitation of the $\De$, the $\De^{1/2-}(1620)$ and $\De^{3/2-}(1700)$ pair, consists of a diquark with  $I_D=S_D=1$ and an anti-diquark with spin and isospin zero. Therefore, the total spin of the tetraquark is $S_T=1$, the isospin is one and $L_T=1$. The parity is $P=-$, and the possible values for $J$ are: $J=0, ~1,~ 2$. Like in the case of the tetraquark superpartner of the $\De$, such a state does not have  definite $C$ but states with definite charge conjugation can be obtained by the symmetric and antisymmetric superposition in Eq.~\req{T1pm}. This leads to the quantum numbers $I~(J^{PC})=1~(0^{-\pm})$,  $1~(1^{-\pm})$ and $1~(2^{-\pm})$ for the tetraquark superpartners of the $\De^{1/2-}(1620)$ and $\De^{3/2-}(1700)$ pair. States with $J^{PC}=0^{-+},~1^{--}$ and $2^{-+}$ will mix with mesonic ($q\bar{q}$) states and will not be considered here as candidates for genuine tetraquark states. On the other hand, states with  $J^{PC}=0^{--}, ~1^{-+}$ and $2^{--}$ are natural candidates for genuine tetraquark states, since they do not obey the additivity property. It is interesting to notice that there is no observed  bosonic states with $J^{PC}=0^{--}$ or $2^{--}$~\cite{PDG}. This could be considered as an indication that these states are too broad or even unbound, as already pointed out in the case of the tetraquark superpartner of the baryonic states $N^{3/2-}(1520)$ and $N^{1/2-}(1535)$. However, the state $\pi_1(1600)$ with $I~(J^{PC})=1~(1^{-+})$ and mass $(1662\pm10)$ MeV, is an excelent  candidate for the tetraquark superpartner of the $\De^{1/2-}(1620)$ and $\De^{3/2-}(1700)$ pair. It is important to notice that the quantum numbers $J^{PC}=1^{-+}$ are considered exotic. A quark-antiquark state with $P=-$ must have $L$ even, and since  $C=(-1)^{L+S}$, $S$ must also be even if $C=+$. Therefore, $S=0$ and $J=L$ which excludes $J=1$. However, hybrid mesons with explicit gluon,  and tetraquark states can carry such quantum numbers. In particular, in Ref.~\cite{Chen:2008qw} it was shown that the $\pi_1(1600)$ can be interpreted as a tetraquark state with a diquark-antidiquark structure.  Here the interpretation of the $\pi_1(1600)$ as a tetraquark state is natural.

Since, as discussed above,  mesonic states with $I~(J^{PC})=1~(2^{-+})$ will mix with tetraquark states, we will consider only  the $\eta_2(1645)$, and not the $\pi_2(1670)$, as the mesonic superpartner of the $N^{3/2-}(1520)$ and $N^{1/2-}(1535)$ pair.

{\renewcommand{\arraystretch}{1.1}  
\begin{table}[h]
\begin{center}
\begin{tabular}{| ccc | ccc|ccc|}
\hline
\multicolumn{3}{|c|}{Meson} & \multicolumn{3}{c|}{Baryon} & \multicolumn{3}{c|}{Tetraquark}\\
$q$-cont&$J^{P(C)}$ & Name & $q$-cont & $J^{P}$ &Name &$q$-cont &$J^{P(C)}$& Name \\
\hline
$\bar{q}q$&$0^{-+}$&$\pi(140)$& --- &---&---& --- & --- &--- \\
$\bar{q}q$&$1^{+-}$&$h_1(1170)$& $[ud]q$ & $(1/2)^+$&$N(940)$ & $[ud][\bar{u}\bar{d}]$ & $0^{++}$ &$\sigma(500)$ \\
$\bar{q}q$&$2^{-+}$&$\eta_2(1645)$ & $[ud]q$ &$(3/2)^-$&$N_{{3\over2}^-}(1520)$ & $[ud][\bar{u}\bar{d}]$ & $1^{-+}$ & --- \\
\hline
$\bar{q}q$&$1^{--}$&$\rho(770),~\omega(780)$& --- &---&---& --- & --- &--- \\
$\bar{q}q$&$2^{++}$&$a_2(1320),~f_2(1270)$& $(qq)q$ & $(3/2)^+$&$\Delta(1232)$ & $(qq)[\bar{u}\bar{d}]$ & $1^{++}$ &$a_1(1260)$ \\
$\bar{q}q$&$3^{--}$&$\rho_3(1690),~\omega_3(1670)$ & $(qq)q$ & $(3/2)^-$&$\Delta_{{3\over2}^-}(1700)$ & $(qq)[\bar{u}\bar{d}]$ & $1^{-+}$ & $\pi_1(1600)$ \\
  $\bar{q}q$&$4^{++}$&$a_4(2040),~f_4(2050)$ & $(qq)q$ & $(7/2)^+$ & $\Delta_{{7\over2}^+}(1950)$ & $(qq)[\bar{u}\bar{d}]$ & --- & --- \\
\hline
$\bar{q}s$&$0^{-}$&$\bar{K}(495)$& --- &---&---& --- & --- &--- \\
$\bar{q}s$&$1^{+}$&$\bar{K}_1(1270)$& $[ud]s$ & $(1/2)^+$&$\Lambda(1115)$ & $[ud][\bar{s}\bar{q}]$ & $0^+$ &$K_0^*(1430)$ \\
$\bar{q}s$&$2^{-}$&$K_2(1770)$ & $[ud]s$ & $(3/2)^-$&$\Lambda(1520)$ & $[ud][\bar{s}\bar{q}]$ & $1^{-}$ & --- \\
\hline
$\bar{s}q$&$0^{-}$&${K}(495)$& --- &---&---& --- & --- &--- \\
$\bar{s}q$&$1^{+}$&$K_1(1270)$& $[sq]q$ & $(1/2)^+$&$\Sigma(1190)$ & $[sq][\bar{s}\bar{q}]$ & $0^{++}$ &$a_0(980)$ \\
&&&&&&&&$f_0(980)$ \\
\hline
$\bar{s}q$&$1^{-}$&$K^*(890)$& --- &---&---& --- & --- &--- \\
$\bar{s}q$&$2^{+}$&$K_2^*(1430)$& $(sq)q$ & $(3/2)^+$&$\Sigma(1385)$ & $(sq)[\bar{u}\bar{d}]$ & $1^{+}$ &${K}_1(1400)$ \\
$\bar{s}q$&$3^{-}$&$K_3^*(1780)$ & $(sq)q$ & $(3/2)^-$ & $\Sigma(1670)$ & $(sq)[\bar{u}\bar{d}]$ & $2^{-}$ & $K_2(1820)$ \\
$\bar{s}q$&$4^{+}$&$K_4^*(2045)$ & $(sq)q$ & $(7/2)^+$ & $\Sigma(2030)$ & $(sq)[\bar{u}\bar{d}]$ & --- & --- \\
 \hline
$\bar{s}s$&$0^{-+}$&$\eta^\prime(958)$& --- &---&---& --- & --- &--- \\
 $\bar{s}s$&$1^{+-}$&$h_1(1380)$& $[sq]s$ & $(1/2)^+$&$\Xi(1320)$ & $[sq][\bar{s}\bar{q}]$ & $0^{++}$ &$f_0(1370)$ \\
 &&&&&&&&$a_0(1450)$\\
$\bar{s}s$&$2^{-+}$&$\eta_2(1870)$ & $[sq]s$ & $(3/2)^-$ & $\Xi(1620)$ & $[sq][\bar{s}\bar{q}]$ & $1^{-+}$ & --- \\
\hline
$\bar{s}s$&$1^{--}$&$\Phi(1020)$& --- &---&---& --- & --- &--- \\
 $\bar{s}s$&$2^{++}$&$f_2^\prime(1525)$& $(sq)s$ & $(3/2)^+$&$\Xi^*(1530)$ & $(sq)[\bar{s}\bar{q}]$ & $1^{++}$ &$f_1(1420)$ \\
&&& & &&&&$a_1(1420)$ \\
$\bar{s}s$&$3^{--}$&$\Phi_3(1850)$ & $(sq)s$ & $(3/2)^-$ & $\Xi(1820)$ & $(sq)[\bar{s}\bar{q}]$ & --- & --- \\
\hline
 $\bar{s}s$&$2^{++}$&$f_2(1640)$& $(ss)s$ & $(3/2)^+$&$\Omega(1672)$ & $(ss)[\bar{s}\bar{q}]$ & $1^{+}$ &$K_1(1650)$ \\
\hline
\end{tabular}
\end{center}
\caption{\small \label{tetratable} Quantum numbers of the states and constituent clusters of different meson families (with only light quarks: $q=u,~d$ and $s$) and their supersymmetric baryon and tetraquark partners. Each family is separated by a horizontal line. For a $q\bar{q}$ state $P=-(-1)^{L_M},~C=(-1)^{L_M+S_M}$.
  For the baryons multiplets with same $L_B$ and $S_D$ we show only the state with the highest possible value for $J$. Diquarks represented by $[~]$ have total spin $S_D=0$, and the ones represented by $(~)$ have $S_D=1$.} 
\end{table}

Using similar analysis, we can  determine the quantum numbers for the tetraquark superpartners, and we can assign a tetraquark candidate to all the baryonic states with $L_B=0$.  Since, as discussed above, tetraquark states can be very broad, we do not make predictions for tetraquark superpartners of baryonic states with $L_B\neq0$ and we only assign a tetraquark superpartner  for these baryons when there is a natural cadidate. Here we consider only the hadronic families with the radial quantum number $n=0$. The superpartners are shown in Table~\ref{tetratable} for the hadrons containing only light quarks, the $u,~d$ and $s$ quarks.
From Table~\ref{tetratable} we can see that it is possible to assign a baryon superpartner to all mesons with $J=L_M+S_M$  and a tetraquark superpartner to all baryons with $L_B=0$, listed in Ref.~\cite{PDG}. 

It is interesting to notice that in Ref.~\cite{Brodsky:2014yha} the states
  $K_1(1400)$ and $K_2(1820)$ were, tentatively, assigned as the $n=1$ partners of the states $K_1(1270)$ and $K_2(1770)$. However, as can be seen in Fig.~(5.2) of  Ref.~\cite{Brodsky:2014yha}, the agreement was poor. Here these two states are interpreted as the tetraquark superpatners of the baryonic states $\Sigma(1385)$ and $\Sigma(1670)$ (as shown in Table~\ref{tetratable}) with a much better agreement.

The case of the state $\Xi^*(1530)$ is intriguing. Naively one could expect such a state to be a $|(ss)q\rangle$ state. However, in the LFHQCD approach its quark structure is $|(sq)s\rangle$, since the supercharge transforms the antiquark $\bar{s}$ in the meson into a diquark cluster $(sq)$. The tetraquark superpartner of a $|(ss)q\rangle$ state would have the structure $|(ss)[\bar{u}\bar{d}]\rangle$ and, therefore, total strangeness $-2$. However, a $|(sq)s\rangle$ state is the superpartner of the tetraquark $|(sq)[\bar{s}\bar{q}]\rangle$ with total strangeness $0$. This feature is most welcome, since no  bosonic hadrons with total strangeness $-2$ have been observed. Since the isospin of the diquark and the anti-diquark in the tetraquark state is 1/2, tetraquarks with both $I=0$ and $1$ are allowed in this case, as in the case of the $\Sigma(1190)$ and $\Xi(1320)$. Again, a $|(sq)[\bar{s}\bar{q}]\rangle$ is not an eigenstate of $C$ neither of $I$. Eigenstates of $C$ can be obtained by the symmetric and antisymmetric superposition of the states:
\beqa \label{Tqpm}
|T_q^+\rangle &=&{1\over\sqrt{2}}\left(|(sq)[\bar{s}\bar{q}]\rangle+|[sq](\bar{s}\bar{q})\right)\rangle,\;\;J^{PC}=1^{++},\nonumber\\
|T_q^-\rangle &=&{1\over\sqrt{2}}\left(|(sq)[\bar{s}\bar{q}]\rangle-|[sq](\bar{s}\bar{q})\rangle\right),\;\;J^{PC}=1^{+-},
\enqa
and eigenstates of the isospin can be obtained by the superposition of $T_u^\pm$ and $|T_d^\pm\rangle$:
\beq\label{TI}
|T_{I=0}^\pm\rangle={|T_u^\pm\rangle +|T_d^\pm\rangle\over\sqrt{2}},\;\;|T_{I=1}^\pm\rangle={|T_u^\pm\rangle -|T_d^\pm\rangle\over\sqrt{2}}.
\enq
Therefore, the $J^{PC}=1^{++}$ states $f_1(1420)~(I=0)$ and $a_1(1420)~(I=1)$ are candidates. It is very reassuring to notice that if $f_1(1420)$ were just a normal $I=0,~s\bar{s}$ mesonic state it would not have a isovector partner like the $a_1(1420)$. Therefore, as in the case of the states $f_0(980),~a_0(980)$, the mass degeneracy between the states $f_1(1420)$ and $a_1(1420)$ is natural in the  four-quark scenario. We also notice that the $I~(J^{PC})=0~(1^{+-})$ state $h_1(1380)$ does not have an observed isovector partner ``$b_1(1380)$''. We interpret this fact as an indication that for the $J^{PC}=1^{+-}$ quantum numbers the $h_1(1380)$ is a true $s\bar{s}$ state (as considered in Table~\ref{etafa}) and, therefore, does not have an isovector partner.

Regarding the state $\Omega(1672)$, being a $I~(J^P)=0~({3\over2}^+)$ baryonic state, its meson superpartner has to be a $I~(J^{PC})=0~(2^{++})$ and one possible candidate is the $f_2(1640)$, although this state needs confirmation \cite{PDG}. In this case the supercharge operator $Q^\dagger$ transforms the constituent $\bar{s}$ into the diquark $({s}{s})$, and leads to the $\Omega(1672)$. The tetraquark superpartner of the $\Omega$  consists of a diquark with  $I_D=0,\; S_D=1$ and an anti-diquark with isospin 1/2 and spin 0. Therefore, the tetraquark has quantum numbers $I~(J^{P})={1\over2}~(1^{+})$ and quark structure $|(ss)[\bar{s}\bar{q}]\rangle$. We assign the $K_1(1650)$ as the tetraquark superpartner of the $\Omega$.

\subsubsection{Single Heavy Quark States \label{te-Qq}}

{\renewcommand{\arraystretch}{1.1}  
\begin{table}[h]
\begin{center}
\begin{tabular}{| ccc | ccc|ccc|}
\hline
\multicolumn{3}{|c|}{Meson} & \multicolumn{3}{c|}{Baryon} & \multicolumn{3}{c|}{Tetraquark}\\
$q$-cont&$J^{P(C)}$ & Name & $q$-cont & $J^{P}$ &Name &$q$-cont &$J^{P(C)}$& Name \\
\hline
$\bar{q}c$&$0^{-}$&$D(1870)$& --- &---&---& --- & --- &--- \\
$\bar{q}c$&$1^{+}$&${D}_1(2420)$& $[ud]c$ & $(1/2)^+$&$\Lambda_c(2290)$ & $[ud][\bar{c}\bar{q}]$ & $0^{+}$ &$\bar{D}_0^*(2400)$ \\
$\bar{q}c$&$2^{-}$&$D_J(2600)$ & $[ud]c$ & $(3/2)^-$ & $\Lambda_c(2625)$ & $[ud][\bar{c}\bar{q}]$ & $1^{-}$ & --- \\
\hline
$\bar{c}q$&$0^{-}$&$\bar{D}(1870)$& --- &---&---& --- & --- &--- \\
$\bar{c}q$&$1^{+}$&$\bar{D}_1(2420)$& $[cq]q$ & $(1/2)^+$&$\Sigma_c(2455)$ & $[cq][\bar{u}\bar{d}]$ & $0^{+}$ &${D}_0^*(2400)$ \\
\hline
$\bar{q}c$&$1^{-}$&$D^*(2010)$& --- &---&---& --- & --- &--- \\
$\bar{q}c$&$2^{+}$&$D_2^*(2460)$& $(qq)c$ & $(3/2)^+$&$\Sigma_c^*(2520)$ & $(qq)[\bar{c}\bar{q}]$ & $1^{+}$ &${D}(2550)$ \\
$\bar{q}c$&$3^{-}$&$D_3^*(2750)$ & $(qq)c$ & $(3/2)^-$ & $\Sigma_c(2800)$ & $(qq)[\bar{c}\bar{q}]$ & --- & --- \\
 \hline
$\bar{s}c$&$0^{-}$&$D_s(1968)$& --- &---&---& --- & --- &--- \\
$\bar{s}c$&$1^{+}$&${D}_{s1}(2460)$& $[sq]c$ & $(1/2)^+$&$\Xi_c(2470)$ & $[sq][\bar{c}\bar{q}]$ & $0^{+}$ &$\bar{D}_{s0}^*(2317)$ \\
$\bar{s}c$&$2^{-}$&$D_{s2}(\sim2830)$? & $[sq]c$ & $(3/2)^-$ & $\Xi_c(2815)$ & $[sq][\bar{c}\bar{q}]$ & $1^{-}$ & --- \\
\hline
$\bar{s}c$&$1^{-}$&$D_s^*(2110)$& --- &---&---& --- & --- &--- \\
$\bar{s}c$&$2^{+}$&$D_{s2}^*(2573)$& $(sq)c$ & $(3/2)^+$&$\Xi_c^*(2645)$ & $(sq)[\bar{c}\bar{q}]$ & $1^{+}$ &${D}_{s1}(2536)$ \\
$\bar{s}c$&$3^{-}$&$D_{s3}^*(2860)$ & $(sq)c$ & $(1/2)^-$ & $\Xi_c(2930)$ & $(sq)[\bar{c}\bar{q}]$ & --- & --- \\
\hline
$\bar{c}s$&$1^{+}$&$\bar{D}_{s1}(\sim2700)$?& $[cs]s$ & $(1/2)^+$&$\Omega_c(2695)$ & $[cs][\bar{s}\bar{q}]$ & $0^{+}$ & ?? \\
\hline
$\bar{s}c$&$2^{+}$&$D_{s2}^*(\sim2750)$?& $(ss)c$ & $(3/2)^+$&$\Omega_c(2770)$ & $(ss)[\bar{c}\bar{s}]$ & $1^{+}$ & ?? \\
\hline
\end{tabular}
\end{center}
\caption{\small \label{tetratable2}
  Same as Table~\ref{tetratable} but for mesons containing one charm quark. We assign the quantum numbers $J^P=1^+$ and $J^P=2^-$ respectively to the states $D(2550)$ and $D_J(2600)$, but their quantum numbers have not yet been determined. States with a question mark (?) are the predicted ones.}
\end{table}

In Table~\ref{tetratable2} we show the superpartners of hadrons containing one $c$ quark. From this Table we can see that it is possible to assign a baryonic superpartner to all mesons with $J=L_M+S_M$ and $n=0$ listed in Ref.~\cite{PDG}. However, to explain all the observed baryons we had to make predictions for some mesonic states like the meson superpartners of the states $\Omega_c(2695)$ and $\Omega_c(2770)$. Similar to the case of $\Omega(1672)$, here the supercharge operator $Q^\dagger$ transforms the constituent $\bar{c}$, in the $\bar{c}s$ meson, into the diquark $[{c}{s}]$, and this leads to the $\Omega_c(2270)$. On the other hand, $Q^\dagger$ transforms the constituent $\bar{s}$, in the $\bar{s}c$ meson, into the diquark $({s}{s})$, that leads to  the $\Omega_c(2770)$.  Since we are already predicting the mesonic superpartners of the $\Omega_c(2695)$ and $\Omega_c(2770)$ baryonic states, we do not make predictions for their tetraquark superpartners. The prediction for the mass of the state $D_{s2}(2830)$ was done in \cite{Dosch:2016zdv}.

Although the  quantum numbers of the states $\Sigma_c(2800)$ and $\Xi_c(2930)$ have not yet been determined, we identify these states as the first $L$ excitation of the $\Sigma_c^*(2520)$ and $\Xi_c^*(2645)$ respectively. In the case of the states $D(2550)$ and $D_J(2600)$, their quantum numbers have not yet been determined either. Based on their decay modes:  $D(2550)\to D^*\pi$ and $D_J(2600)\to D\pi, ~D^*\pi$, we identify them as the $J^P=1^+$ and $2^-$ states respectively. 

{\renewcommand{\arraystretch}{1.1}  
\begin{table}[h]
\begin{center}
\begin{tabular}{| ccc | ccc|ccc|}
\hline
\multicolumn{3}{|c|}{Meson} & \multicolumn{3}{c|}{Baryon} & \multicolumn{3}{c|}{Tetraquark}\\
$q$-cont&$J^{P(C)}$ & Name & $q$-cont & $J^{P}$ &Name &$q$-cont &$J^{P(C)}$& Name \\
\hline
$\bar{q}b$&$0^-$&$\bar{B}(5280)$& --- &---&---& --- & --- &--- \\
$\bar{q}b$&$1^+$&$\bar{B}_1(5720)$& $[ud]b$ & $(1/2)^+$&$\Lambda_b(5620)$ & $[ud][\bar{b}\bar{q}]$ & $0^{+}$ &${B}_J(5732)$ \\
$\bar{q}b$&$2^{-}$&$\bar{B}_J(5970)$ & $[ud]b$ & $(3/2)^-$ & $\Lambda_b(5920)$ & $[ud][\bar{b}\bar{q}]$ &$1^{-}$ & --- \\
\hline
$\bar{b}q$&$0^{-}$&${B}(5280)$& --- &---&---& --- & --- &--- \\
$\bar{b}q$&$1^{+}$&${B}_1(5720)$& $[bq]q$ & $(1/2)^+$&$\Sigma_b(5815)$ & $[bq][\bar{u}\bar{d}]$ & $0^{+}$ &$\bar{B}_J(5732)$ \\
\hline
$\bar{q}b$&$1^{-}$&$B^*(5325)$& --- &---&---& --- & --- &--- \\
$\bar{q}b$&$2^{+}$&$B_2^*(5747)$& $(qq)b$ & $(3/2)^+$&$\Sigma_b^*(5835)$ & $(qq)[\bar{b}\bar{q}]$ & $1^{+}$ &${B}_J(5840)$ \\
  \hline
$\bar{s}b$&$0^{-}$&$B_s(5365)$& --- &---&---& --- & --- &--- \\
$\bar{s}b$&$1^{+}$&${B}_{s1}(5830)$& $[qs]b$ & $(1/2)^+$&$\Xi_b(5790)$ & $[qs][\bar{b}\bar{q}]$ & $0^{+}$ &$\bar{B}_{s0}^*(\sim5800)$? \\
\hline
$\bar{s}b$&$1^{-}$&$B_s^*(5415)$& --- &---&---& --- & --- &--- \\
$\bar{s}b$&$2^{+}$&$B_{s2}^*(5840)$& $(sq)b$ & $(3/2)^+$&$\Xi_b^*(5950)$ & $(sq)[\bar{b}\bar{q}]$ & $1^{+}$ &${B}_{s1}(\sim5900)$? \\
\hline
$\bar{b}s$&$1^{+}$&${B}_{s1}(\sim6000)$?& $[bs]s$ & $(1/2)^+$&$\Omega_b(6045)$ & $[bs][\bar{s}\bar{q}]$ & $0^{+}$ & ?? \\
\hline
\end{tabular}
\end{center}
\caption{\small \label{tetratable3}
  Same as Table~\ref{tetratable} but for mesons containing botton quarks. We assign the quantum numbers $J^P=1^+,~J^P=0^+ $ and $J^P=2^-$ respectively to the states $B_J(5732)$, $B_{J}^*(5840)$ and $B_J(5970)$, but their quantum numbers have not yet been determined. States with a question mark (?) are the predicted ones.}
\end{table}

In Table~\ref{tetratable3} we show the superpartners of hadrons containing one $b$ quark. As in the previous cases it is possible to assign a baryonic superpartner to all mesons with  one $b$ quark and with $J=L_M+S_M$ and $n=0$, listed in Ref.~\cite{PDG}. The only prediction we had to made for the mesons was for the superpartner of the $\Omega_b(6045)$ state and, as in the case of the $\Omega_c(2695)$, we do not make prediction for its tetraquark superpartner. Using the prediction done in \cite{Dosch:2016zdv} we have identified the state $B_J(5970)$ as the $L_M=2,~S_M=0$ mesonic state. We have also assigned the quantum number $J^P=1^+$ and $J^P=0^+$ respectively to the states $B_J(5732)$ and $B_{J}^*(5840)$ whose quantum numbers have not yet been determined. With these assignments, the only candidates for tetraquark states that have not yet been observed are the predicted scalar $B_{s0}^*(\sim5800)$ and axial $B_{s1}(\sim5900)$, as can be seen from Table~\ref{tetratable3}.

Regarding the state $B_{s0}^*$ our prediction is a mass around 5800 GeV. In 2016
the D0 collaboration reported the observation of a narrow structure, called
$X^\pm(5568)$, in the decay $X^\pm(5568)\to B_{s0}\pi^\pm$ \cite{D0:2016mwd}. This state would have a $(us\bar{b}\bar{d})$ quark content  and $I~(J^P)=1~(0^+)$ quantum numbers, as the $B_{s0}^*$ predicted here. The existence of such a state was not confirmed by LHCb Collaboration \cite{Aaij:2016iev}, neither by the CMS Collaboration \cite{CMS-X}. As a matter of fact, the LHCb Collaboration sees no structure in the $B_{s}^0\pi^\pm$ mass spectrum from the $B_{s}^0\pi^+$ threshold up to $M_{B_{s}^0\pi^+}\leq$ 5700 GeV, but there are no restrictions for higher masses. In 2017 the D0 collaboration has confirmed the observation of this state with a significance of $6.7\sigma$~\cite{Abazov:2017poh}.  Although our prediction is $\sim 5800$ MeV, considering the variation in the masses of the superpartner states observed in the other cases, we can not exclude a mass as the one reported by D0 collaboration. A prediction for a diquark-antidiquark tetraquark state, $B_{s0}^*$, with higher mass was also done in \cite{Zanetti:2016wjn}.

\subsubsection{Double Heavy Quark States \label{te-QQ}}

The LF confinement potential for systems containing two heavy quarks will evidently be modified, see, e.g., Ref.~\cite{Trawinski:2014msa}.  A system consisting of two light quarks, or one light and one heavy quark, is relativistic. However, a system consisting of two heavy quarks is close to the non-relativistic case. Therefore, the extension of superconformal algebra to such states is somewhat speculative, and the statements made for the case of states with two heavy quarks should be taken as propositions worth testing. 
However, as we shall show, the excellent agreement between the masses of the superpartners in the double-heavy-quark sector supports this attempt.

{\renewcommand{\arraystretch}{1.1}  
\begin{table}[h]
\begin{center}
\begin{tabular}{| ccc | ccc|ccc|}
\hline
\multicolumn{3}{|c|}{Meson} & \multicolumn{3}{c|}{Baryon} & \multicolumn{3}{c|}{Tetraquark}\\
$q$-cont&$J^{P(C)}$ & Name & $q$-cont & $J^{P}$ &Name &$q$-cont &$J^{P(C)}$& Name \\
\hline
$\bar{c}c$&$0^{-+}$&$\eta_c(2984)$& --- &---&---& --- & --- &--- \\
 $c\bar{c}$&$1^{+-}$&$h_c(3525)$& $[cq]c$ & $(1/2)^+$&$\Xi_{cc}^{SELEX}(3520)$ & $[cq][\bar{c}\bar{q}]$ & $0^{++}$ &$\chi_{c0}(3415)$ \\
&&&&&$\Xi_{cc}^{LHCb}(3620)$ &&&\\
\hline
$\bar{c}c$&$1^{--}$&$J/\psi(3096)$& --- &---&---& --- & --- &--- \\
 $\bar{c}c$&$2^{++}$&$\chi_{c2}(3556)$& $(cq)c$ & $(3/2)^+$&$\Xi_{cc}^{LHCb}(3620)$ & $(cq)[\bar{c}\bar{q}]$ & $1^{++}$ &$\chi_{c1}(3510)$ \\
\hline
&&n=1&&&&&&\\
$\bar{c}c$&$1^{--}$&$\psi^\prime(3686)$& --- &---&---& --- & --- &--- \\
 $\bar{c}c$&$2^{++}$&$\chi_{c2}(3927)$& $(cq)c$ & $(3/2)^+$&$\Xi_{cc}^*(\sim3900)$? & $(cq)[\bar{c}\bar{q}]$ & $1^{++}$ &$X(3872)$ \\
& &&&&& & $1^{+-}$ &$Z_c(3900)$ \\
\hline
\hline
$\bar{b}b$&$0^{-+}$&$\eta_b(9400)$& --- &---&---& --- & --- &--- \\
 $b\bar{b}$&$1^{+-}$&$h_b(9900)$& $[bq]b$ & $(1/2)^+$&$\Xi_{bb}(\sim9900)$? & $[bq][\bar{b}\bar{q}]$ & $0^{++}$ &$\chi_{b0}(9860)$ \\
\hline
$\bar{b}b$&$1^{--}$&$\Upsilon(9460)$& --- &---&---& --- & --- &--- \\
 $\bar{b}b$&$2^{++}$&$\chi_{b2}(9910)$& $(bq)b$ & $(3/2)^+$&$\Xi_{bb}(\sim9900)$? & $(bq)[\bar{b}\bar{q}]$ & $1^{++}$ &$\chi_{b1}(9893)$ \\
\hline
&&n=1&&&&&&\\
$\bar{b}b$&$1^{--}$&$\Upsilon(2S)(10020)$& --- &---&---& --- & --- &--- \\
 $\bar{b}b$&$2^{++}$&$\chi_{b2}(10270)$& $(bq)b$ & $(3/2)^+$&$\Xi_{bb}(\sim10300)$? & $(bq)[\bar{b}\bar{q}]$ & $1^{++}$ &$X_{b}^{QCDSR}(10250)$? \\
\hline
\end{tabular}
\end{center}
\caption{\small \label{tetratable4}
  Same as Table~\ref{tetratable} but for mesons containing two heavy quarks. States with a question mark (?) are the predicted ones.}
\end{table}

In Table~\ref{tetratable4} we show the superpartners of hadrons containing two heavy quarks. The baryonic superpartner of the meson $h_c(3525)$ is the $\Xi_{cc}$ state with quantum numbers $J^{P}={1\over2}^+$. There are two candidates for this state:
the $\Xi_{cc}^{SELEX}(3520)$ state observed by the SELEX collaboration in 2002 \cite{Mattson:2002vu,Ocherashvili:2004hi}, and the $\Xi_{cc}^{LHCb}(3620)$ observed in 2017 by the LHCb collaboration \cite{Aaij:2017ueg}. The mass observed by the LHCb collaboration is very close to the prediction made a few years ago in Ref.~\cite{Karliner:2014gca}. If there are two  $\Xi_{cc}$ states or if there is only one is still in debate. However, because the production environments of these two experiments differ from each other, the LHCb collaboration did not exclude the original observations \cite{Aaij:2017ueg}. Since the mass of the SELEX observation is closer to the $h_c(3525)$ mass than LHCb observation, and since the quantum numbers for the $\Xi_{cc}$ state were not determined, here we include the state observed  by the LHCb collaboration with two possible $J^P$ assignments, as shown in Table~\ref{tetratable4}. For more interpretations about these states see Ref.~\cite{Brodsky:2017ntu}. For the two $\Xi_{cc}$ states, $J^{PC}={1\over2}^+$ and $J^{PC}={3\over2}^+$ there are observed tetraquarks superpartners candidates also included in Table~\ref{tetratable4}.

To assign the tretraquark partner to states with two heavy quarks we follow here the standard procedure described in Sec.~\ref{te}, {\it i.e.}, the operator $Q^\dagger$ transforms the active heavy quark (the heavy quark that it is not in the diquark cluster, see Fig.~\ref{tetra2}) in the baryon into an heavy-light anti-diquark cluster with spin zero. We refer the reader to Ref.~\cite{Dosch-lec} for another possibility.

The case of the $\chi_{c2}(3556)$ and $\chi_{c1}(3510)$ meson-tetraquark superpartners deserves special attention. As discussed above in the strange sector,
the diquark structure of the tetraquark superpartner of the state $\Xi_{cc}^*$, $J^P=(3/2)^+$ is $(cq)[\bar{c}\bar{q}]$, which is not an eigenstate of $C$ neither of $I$. Eigenstates of $C$ and $I$ can be obtained by the symmetric and antisymmetric superposition of the states as shown in Eqs.~\req{Tqpm} and \req{TI} (exchanging $s\to c$). Therefore, the $\chi_{c1}(3510), ~1^{++}$ state is the natural tetraquark candidate with $I=0$, and the $h_c(3525), ~1^{+-}$ state can be a mixture between $c\bar{c}$ and a $I=0$ tetraquark state. No $I=1$ partners of $\chi_{c1}(3510)$ and $h_c(3525)$ were observed and this could be interpreted as an indication that $h_c(3525)$ is just a $c\bar{c}$ state and that the isovector partner of  $\chi_{c1}(3510)$ is unbound or to broad to be observed. However, the case of the $n=1$ (see Eqs.\req{mesfin} and \req{tetrafin}) partners of the  $h_c(3525)$ and  $\chi_{c1}(3510)$ families, the $\chi_{c2}(3927)$ and $X(3872)$ states, is different.

The $X(3872)$ is the most well studied among the new charmonium states observed in the last years. It was first observed  in 2003 by  Belle 
Collaboration \cite{Choi:2003ue,Adachi:2008te},  and has been confirmed by  
five collaborations: BaBar~\cite{Aubert:2008gu}, CDF~
\cite{Acosta:2003zx,Abulencia:2006ma,Aaltonen:2009vj}, 
\DZero~\cite{Abazov:2004kp}, LHCb~\cite{Aaij:2011sn,Aaij:2013zoa}
and  CMS~\cite{Chatrchyan:2013cld}. 
The LHCb collaboration has determined the $X(3872)$ quantum numbers to be $J^{PC} = 1^{++}$, with more than 8$\sigma$ significance \cite{Aaij:2013zoa}. Calculations using constituent quark models give masses for  possible charmonium states, with $J^{PC}=1^{++}$ quantum numbers,  which are much bigger than the observed $X(3872)$ mass: $2~^3P_1(3990)$ and $3~^3P_1(4290)$ \cite{bg}. These results, together with the coincidence between the $X$ mass and the $D^{*0}D^0$ threshold: $M(D^{*0}D^0)=(3871.81\pm0.36)$MeV \cite{PDG}, inspired the proposal that the $X(3872)$ could be a molecular $(D^{*0}\bar{D}^0+\bar{D}^{*0}D^0)$ bound state with a small binding energy \cite{swanson}. Other interesting possible interpretation of the $X(3872)$, first proposed in Ref.~\cite{maiani}, is that it could be a tetraquark state resulting from the binding of a diquark and an antidiquark. In the LFHQCD approach the interpretation of the $X(3872)$ as a tetraquark state is natural, and this could help to settle the controversy about its structure.

In Ref.~\cite{maiani} a $I~(J^{PC})=1~(1^{+-})$ tetraquark state was predicted with a mass close to the $X(3872)$ mass, and such a state was indeed observed in 2015 by BESIII  and Belle Collaborations, the $Z_c^\pm(3900)$ \cite{Ablikim:2013mio,Liu:2013dau}. The existence of this structure was promptly confirmed by the  authors of Ref.~\cite{Xiao:2013iha} using  CLEO-c data. In the LFHQCD approach the existence of a $I~(J^{PC})=1~(1^{+-})$ charmonium tetraquark state is also predicted. Clearly the baryonic superpartner of the $\chi_{c2}(3927)$ and $X(3872)$ states is a $n=1$, $J^P={3/2}^+$, $\Xi_{cc}$ state whose mass we predict to be around 3900 GeV. For completeness, we include this particular case of states with $n=1$ in Table~\ref{tetratable4}.

Another state between the new charmonium states that can also be considered as a tetraquark candidate is the $X(3915)$. It was first observed by  {\sc Belle} Collaboration in  the decay $B\to  (J/\psi\,\omega)K $, with a slightly higher mass: $m=3943\pm11({\mbox{{\it stat}}})\pm13({\mbox{{\it syst}}})$ MeV \cite{belleY3940}. Soon after, this state has been also observed  in the process $B\to  (J/\psi\,\omega)K$  by  {\sc Babar} Collaboration,     with     a mass  $m=3914.6^{+3.9}_{-3.4}~({\mbox{{\it stat}}})~\pm~2.0({\mbox{\it syst}})$   MeV    and   width   $\Gamma=34^{+12}_{-8}~({\mbox{{\it stat}}})~\pm~5.0({\mbox{\it  syst}})$ MeV  \cite{babarY3940}. The $X(3915)$ state has a larger product of the two-photon width times the decay branching fraction than usually expected for  charmonium states, as noted in Ref. \cite{babar2}.  Although the Particle Data Group \cite{PDG} assigns the label $X(3915)$ to the  charmonium state $\chi_{c0}(3915)$, it does allow the possible  quantum numbers $J^{PC}=0^{++}$ or $2^{++}$. Considering that the state $\chi_{c2}(3927)$ has a mass very close to $X(3915)$, it seems more likely to associate $J^{PC}=0^{++}$ as the $X(3915)$ quantum numbers. In  this case, this state should be the analog of the $X(3872)$ in the $n=1$ family of $\eta_c(2984)$ and it could be associated as the tetraquark superpartner of the predicted $h_c(2P)$ state \cite{Lebed:2016hpi}, the $n=1$ partner of the $h_c(3525)$.

These two examples show how the LFHQCD approach can be used to help clarify the structure of the new charmonium states.

For the states with two $b$ quarks, although baryonic states with double $b$ quarks have not yet been observed, we can still identify a clear candidate for the tetraquark superpartner of the meson $h_b(9900)$, the scalar state $\chi_{b0}(9860)$. The same for the meson-tetraquark superpartners $\chi_{b2}(9910)$ and $\chi_{b1}(9893)$. As was done for the $c\bar{c}$ case, we have also included in Table~\ref{tetratable4} the n=1 $\chi_{b2}(10270)$ state, from the $\Upsilon(2S)$ family, since its tetraquark superpartner would be the b-state analog to the $X(3872)$. This state was predicted in Ref.~\cite{Matheus:2006xi} from a QCD sum rule (QCDSR) study of diquark-antidiquark states.

\section{Conclusions}

Supersymmetric quantum mechanics together with light-front holography can account for  principal features of hadron physics, such as the approximatively linear Regge trajectories  with nearly equal slopes for all mesons, baryons and tetraquarks in both $L$  and $n$. We emphasize that the supersymmetric features of hadron physics derived from superconformal quantum mechanics refers to the symmetry properties of  the bound-state wave functions of hadrons and not to quantum fields; there is therefore no need to introduce new supersymmetric  fields or particles such as squarks or gluinos. The  tetraquarks are required to complete the supermultiplets predicted by the superconformal algebra \cite{Brodsky:2016yod}. The tetraquarks are the bound states of the same confined color-triplet diquarks and anti-diquarks which account for baryon spectroscopy.

The masses of mesons and their superpartner baryons are related by $L_M=L_B+1$ and $S_M=S_D$, and the masses of baryons and their superpartner tetraquarks are related by $L_B=L_T$ and $S_D=S_T$.  As can be seen by Tables~\ref{tetratable}, \ref{tetratable2} and  \ref{tetratable3}, the agreement with experiment is generally better for the trajectories with total diquark spin $S_D=1$, such as the $\rho-\Delta-a_1$ superpartner trajectories,  than for the $\pi-N-\sigma$ trajectories ( $S_D=0$). The model incorporates features expected from  chiral symmetry, such as a masslessness pion in the massless quark limit. The structure of the superconformal algebra also implies that the pion and the $\rho$ have no supersymmetric partner. The meson-baryon-tetraquark supersymmetry is observed to survive even if the heavy quark masses strongly break the conformal symmetry~\cite{Dosch:2015bca,Dosch:2016zdv}. In this approach, the heavy quark influences the spectrum indirectly by modifying the strength of the harmonic potential; this modification cannot be determined from supersymmetry. However, as shown in Ref.~\cite{Dosch:2016zdv}, the dependence of the confinement scale on the heavy quark mass can be calculated in HQET, and it is in agreement with the observed increase.

The supercharge operator, $Q^\dagger$, transforms a quark into a two-antiquark cluster. In the standard procedure, allowed by the 4-supermultiplet structure followed here, the operator $Q^\dagger$ acts only in the quark not in the diquark cluster. Therefore, it is not possible to obtain states with more than four constituents. This is a very robust prediction from the model and can help to answer the question posed in Ref.~\cite{Selem:2006nd}: why multi-nucleons do not  merge into a single color singlet cluster, e.g. $qqqqqq$. From this LFHQCD approach we conclude that all possible states with more than 4 constituents are necessarily molecular states, like the deuteron and all nuclei.  Therefore, one can also conclude that pentaquark states, if they exist, have to be molecular states, and not compact states with two diquarks and one quark. This can be related with the ``repulsive force between diquarks'' proposed in \cite{Selem:2006nd}.

There are other two very robust predictions that can be made from this  LFHQCD approach: i) all scalar ($J^{PC}=0^{++}$) and axial-vector ($J^{PC}=1^{++}$) states, as well as bosonic hadrons with exotic quantum numbers (as for instance $J^{PC}=1^{-+}$)  are in fact tetraquark states; ii) baryonic bound states with three $c$ (or $b$) quarks are disfavored. This last prediction follows directly from the fact that, due to its large mass, it is unlikely to transform a $\bar{c}$ ($\bar{b}$) antiquark in a $(cc)$ ($(bb)$)  diquark cluster and, therefore, no $(cc)c$ ($(bb)b$) states are predicted.

It is also very interesting to make comparisons with previous attempts to organize the spectrum of the hadrons as done, for example, in Refs.~\cite{Catto:1984wi,Selem:2006nd,Klempt:2009pi,Forkel:2007tz,Forkel:2008un}. In Ref.~\cite{Catto:1984wi} mesons, baryons and tetraquarks are also hadronic states within the same multiplet. The diquark structures proposed for the baryons in Table~\ref{tetratable} are the same as in Ref.~\cite{Selem:2006nd} and the near-degeneracy between the baryon and mesons superpartners was also pointed out in \cite{Selem:2006nd} as a consequence of what the authors call ``near-degeneracy between the good diquark and a light antiquark''. In Ref.~\cite{Klempt:2009pi} the $J^P=(3/2)^-$ assignment was also done for the state $\Sigma_c(2800)$. Linear Regge trajectories for the mass spectrum of light mesons and baryons with $L_B=L_M-1$ were also found in Ref.~\cite{Forkel:2007tz,Forkel:2008un}, using AdS/QCD in the soft-wall approximation.  

Finally, as can be seen from Tables~\ref{tetratable}, \ref{tetratable2} and \ref{tetratable3}, it is possible to associate a baryonic superpartner to all mesons with quantum numbers satisfying the additivity property ($J=L_M+S_M$)  listed in Ref.~\cite{PDG}.  The present approach can also be very useful in identifying the structure of the new charmonium states, as shown, for example, by the identification of the  $X(3872)$ and $X(3915)$ states as $(cq)[\bar{c}\bar{q}]$ tetraquarks. We emphasize that, in contrast to standard diquark models, LFHQCD does not generate an excess set of tetraquark states. Historically, the study of hadron properties has led to some of the most relevant discoveries in particles physics and, thus still remains an invaluable guide to fundamental  features of QCD.

\section*{Acknowledgments}

The authors would like to deeply thank H.G. Dosch for clarifying discussions and suggestions. Many assignments described here would not be possible whithou his valuable contribution. The authors would also like to thank G. de T\'eramond for all the help and valuable discussions.
S.J.B. is supported by the Department of Energy, contract DE--AC02--76SF00515.  M.N is supported by FAPESP process\# 2017/07278-5. SLAC-PUB-17231.

\end{document}